\documentclass[11pt]{article}
\usepackage{amsmath,amssymb,color,epsfig,cite}
\usepackage{graphicx}
\usepackage{subfigure}
\usepackage{setspace}
\usepackage{amsfonts}
\newcommand{\hoch}[1]{$\, ^{#1}$}

\makeatletter
\@addtoreset{equation}{section}
\makeatother

\newcommand{\be}{\begin{equation}}
\newcommand{\ee}{\end{equation}}
\newcommand{\bea}{\setlength\arraycolsep{2pt} \begin{eqnarray}}
\newcommand{\eea}{\end{eqnarray}}
\newcommand{\nn}{\nonumber}

\def\ft#1#2{{\textstyle{\frac{\scriptstyle #1}{\scriptstyle #2} } }}
\def\fft#1#2{{\frac{#1}{#2}}}

\def\0{{\sst{(0)}}}
\def\1{{\sst{(1)}}}
\def\2{{\sst{(2)}}}
\def\3{{\sst{(3)}}}
\def\4{{\sst{(4)}}}
\def\5{{\sst{(5)}}}
\def\6{{\sst{(6)}}}
\def\7{{\sst{(7)}}}
\def\8{{\sst{(8)}}}
\def\9{{\sst{(9)}}}

\def\sst#1{{\scriptscriptstyle #1}}

\def\ep{{\epsilon}}

\begin{document}

%\begin{flushright}
%\hfill{MI-TH-1762}

%\end{flushright}

\begin{center}
{\large {\bf Black Holes That Repel}}

\vspace{10pt}

H. L\"u\hoch{1\dagger}, Zhao-Long Wang\hoch{2,3*} and Qing-Qing Zhao\hoch{1\ddagger}

\vspace{15pt}

\hoch{1}{\it Center for Joint Quantum Studies, School of Science\\
                  Tianjin University, Tianjin 300350, China}

\vspace{10pt}

\hoch{2}{\it Institute of Modern Physics, Northwest University, XiAn 710069, China}

\vspace{10pt}

\hoch{3}{\it Shaanxi Key Laboratory for Theoretical Physics Frontiers, XiAn 710069, China}

\vspace{40pt}

\underline{ABSTRACT}
\end{center}

The recent observation that black holes in certain Einstein-Maxwell-Dilaton (EMD) theories can violate the entropy super-additivity led to the suggestion that these black holes might repel each other. In this paper, we consider EMD theories with two Maxwell fields $A_i$, with general exponential couplings $\exp(a_i \phi)$ in their kinetic terms. We calculate the gravi-electrostatic force between charged black holes $(m_1,e_1)$ and $(M_2,Q_2)$; the former is sufficiently small and can be treated as a point-like object. We find there is a potential barrier caused by the dilaton coupling at $r_0$ outside the back hole horizon $r_+$, provided that $-a_1 a_2> 2(D-3)/(D-2)$. As the black hole approaches extremality, both $r_+$ and $r_0$ vanish, the barrier becomes infinitesimally thin but infinitely high, and the two black holes repel each other in the whole space.  There is no electrostatic force between them; the dilaton is the antigravity agent. Furthermore we find that the exact constraint on $a_1 a_2$ can be derived from the requirements that two-charged extremal black holes have a fission bomb like mass formula and the violation of entropy super-additivity can occur. The two very different approaches give a consistent picture of the black hole repulsion.

\vfill {\footnotesize \hoch{\dagger}mrhonglu@gmail.com\ \ \ \hoch{*}wangzl@nwu.edu.cn\ \ \ \hoch{\ddagger}zhaoqq@tju.edu.cn}

\pagebreak
%\voffset=0pt
%\setcounter{page}{1}

%\tableofcontents
%\addtocontents{toc}{\protect\setcounter{tocdepth}{2}}

%%%%%%%%%%%%%%%%%%%%%%%%%%%%%%%%%%%%%%%%

%\newpage
%%%%%%%%%%%%%%%%%%%%%%%%%%%%%%%%%%%%%%%%

\section{Introduction}
\label{sec:intro}

Newton's law of universal gravitation states that any two massive objects attract by gravity. This remains largely true in Einstein's theory of General Relativity (GR) but with some subtleties. The centrifugal force is viewed as fictitious in Newton's theory, it is an intrinsic effect of spacetime structure in GR. It is thus advantageous to simplify the discussion on the effective force between two objects by restricting the motion with no relative angular momentum.  Furthermore, in de Sitter spacetimes or any universe with accelerating expansion, two black holes in sufficient separation will move away from each other by the negative pressure. In order to avoid this effect, we shall focus only on the black holes that are asymptotic to the Minkowski spacetime.

In Newtonian gravity, the attractive force can be perfectly balanced by the Coulomb's electrostatic force, the other long-range force in nature.  In particular, a set of charged particles of mass $m$ and charge $e$ can all be in static balance when $m=e$. This property remains in GR even when the theory becomes highly nonlinear.  The $m=e$ particles become the extremal Reissner-Nordstr\"om (RN) black holes.  These black holes continue to experience no force  and statically they can be arbitrarily located in space. The $m>e$ particles become non-extremal black holes with Hawking temperature.  The $m<e$ particles, such as an electron, remain particle-like with naked curvature singularity.

There is no doubt that two Schwarzschild black holes always attract; in fact, they can remain static only with a naked strut singularity \cite{Gibbons:1974zd,mulsei}.  The situation becomes more complicated for RN black holes. A direct calculation of the force is formidable.  When one of the two black holes is sufficiently small, it can be treated as a point-like object. It was shown that there can be no static equilibrium for a charged particle with $m>e$ outside the event horizon of an RN black hole \cite{Zhao:2018wkl}.  This is indicative that two RN black holes always attract.  (Hairy magnetic black holes were constructed to lower the mass so that two such black holes can be repulsive at large separation, see, e.g.~\cite{Lee:1991qs,Horne:1992bi,Lee:1994sk}.)
The attractiveness is also supported by the fact that the RN black hole entropy is a super-additive function of its mass and charge.  A black hole's entropy is in general a function of its conserved quantities $Q^i$, which include the mass, charge and angular momentum.  If we imagine that two black holes with $Q_1^i$ and $Q_2^i$ join to become a bigger black hole, entropy super-additivity states $S(Q_1^i + Q_2^i) \ge S(Q_1^i) + S(Q_2^i)$ \cite{Cvetic:2018dqf}. This is related to Hawking's black hole area law and it is consistent with the assumption that black holes are mutually attractive and merging together is a physical process that increases the entropy.  Indeed RN black holes, like the Schwarzschild,  satisfy the entropy super-additivity.

However, it turns out that the super-additivity rule can be violated \cite{Cvetic:2018dqf} by the Kaluza-Klein dyonic black holes in four dimensions \cite{gibwil}. Furthermore, the mass of the extremal dyon is larger than the sum of the masses of the individual extremal electric and magnetic ones. The dyon is thus a bound state of electric and magnetic black holes with negative binding energy, analogous to a fission bomb \cite{gibkal,lupofission}.  The issue of generating angular momentum by the separation of the electric and magnetic charges can be circumvented if we consider two Maxwell fields, with both carrying electric charges.  Further black hole fission bombs in EMD theories that violate the entropy super-additivity were constructed in \cite{Geng:2018jck}.  These results led one to propose that two black holes might not always attract \cite{talks}, and the dilaton may play the role of antigravity \cite{Gibbons:1993dq,gibkal}.

In this paper, we consider EMD theories with two Maxwell fields $A_1$ and $A_2$, both of which couple to the dilation $\phi$ non-minimally, with exponential couplings in their kinetic terms.  We calculate the force on a charged particle coupled to $A_1$ by the black hole carrying only the $A_2$ charge, and hence there is electrostatic force between them.  We find that there is an unstable equilibrium when the black hole becomes extremal or sufficiently near extremum, indicating repulsive force exists between the black hole and the charged particle.  Furthermore, the particle satisfies the black hole mass-charge bound and hence two such black holes do repel each other, with antigravity mediated by the dilaton.

In section 2, we present the formalism for calculating the gravi-electrostatic force.
In sections 3 and 4, we study black hole interactions in EMD theories with one and two Maxwell fields respectively. We conclude the paper in section 5.

\section{Gravi-electrostatic force}

We set up here the formalism for calculating the gravi-electrostatic force between a charged black hole and a point particle outside the horizon. (See e.g.~\cite{Hashimoto:2016dfz}.) We shall consider only the static and spherically-symmetric black holes. The most general ansatz takes the form
\be
ds_D^2 = -h(r) dt^2 + \fft{dr^2}{f(r)} + \rho(r)^2\, d\Omega_{\rm sphere}^2\,,\qquad A=\psi(r)\, dt\,,\qquad \cdots\,,
\ee
where $\psi(r)$ is the electrostatic potential, and
the ellipses denote other matter fields that are not relevant for our purpose.  The motion of a particle of mass $m$ couple to $A$ with charge $e$ is governed by the action
\be
S=-m \int d\tau \Big(\sqrt{-g_{\mu\nu} \ft{dx^\mu}{d\tau} \ft{d x^\nu}{d\tau}} + \ft{e}{m} A_\mu \ft{d x^\mu}{d\tau}\Big)\,,\label{particleaction}
\ee
where $\tau$ is a certain affine parameter.  We are interested in static equilibria; therefore, we focus on the radial motion only.  The relevant action can be written as
\be
S=m \int dt L\,,\qquad L= - \sqrt{h(r) - \ft{\dot r^2}{f(r)}} - \ft{e}{m} \psi(r)\,,
\ee
where the radial variable $r$ is now a function of the asymptotic physical time $t$ and a dot is a derivative with respect to $t$.  To derive the gravi-electrostatic force, we restrict our attention to $\dot r^2\ll 1$, in which case, the effective Lagrangian is
\be
L=\ft{\dot r^2}{2\sqrt{h(r)}\,f(r)} - V_{\rm eff}(r)\,,\qquad V_{\rm eff} = \sqrt{h(r)} + \ft{e}{m} \psi(r)\,.
\ee
This is a Newtonian system with an $r$-dependent mass and potential $V_{\rm eff}$.  When the charge and mass satisfy the ratio at certain $r_0$, namely
\be
\ft{e}{m} = -\ft{(\sqrt{h})'}{\psi'}\big|_{r=r_0}\,,\label{emgen}
\ee
it is a static equilibrium.  The existence of such an equilibrium implies that the particle must experience an overall repulsive force somewhere in its vicinity.  We must require $r_0$ be outside the horizon, and furthermore, the particle mass be sufficiently larger than its charge so that it is a small black hole rather than an electron-like particle. For the linear radial perturbation $r(t)=r_0 + \epsilon(t)$, the solution is $\epsilon\sim \exp(\pm \lambda t)$, with
\be
\lambda^2 =\sqrt{h} f \Big(\ft{\psi''}{\psi'} (\sqrt{h})'-(\sqrt{h})''\Big)\Big|_{r=r_0}\,.
\label{lambdagen}
\ee
The characteristics of the equilibrium is determined by the sign of $\lambda^2$.

It is worth pointing out that in string-inspired EMD theories, there can be multiple Maxwell fields. If the particle of $(m,e)$ does not couple to $A$, the second term in the action (\ref{particleaction}) drops out and there is no electrostatic force.  We then have
\be
h'(r_0)=0\,,\qquad \lambda^2 = -\ft12 f(r_0) h''(r_0)\,.\label{noelec}
\ee

\section{EMD theory}

We first consider the EMD theory in general $D$ dimensions:
\be
{\cal L} = \sqrt{-g} \Big(R - \ft12 (\partial\phi)^2-\ft14 e^{a\phi} F^2\Big)\,,\qquad F=dA\,.
\ee
Introducing $N$ by $a^2=\fft{4}{N} - \fft{2(D-3)}{D-2}$, charged black holes can be written as
\bea
ds^2 &=& -H^{-\fft{D-3}{D-2} N} \tilde f dt^2 +
H^{\fft{N}{D-2}} \Big(\fft{dr^2}{\tilde f} + r^2 d\Omega^2\Big)\,,\quad
A = \psi dt\,,\quad \phi = \ft12 N a \log H\,,\nn\\
\tilde f &=& 1 - \fft{\mu}{r^{D-3}}\,,\qquad \psi = \fft{\sqrt{N\,q(\mu+q)}}{r^{D-3} H}\,,\qquad H=1 + \fft{q}{r^{D-3}}\,.
\label{chargedbh}
\eea
Note that for $a\ne 0$, we have $\rho\rightarrow 0$ as $r\rightarrow 0$; therefore, $r$ can be qualitatively treated as the black hole radius. The horizon is at $r_+=\mu^{1/(D-3)}$, with surface gravity $\kappa = (D-3)/(2r_+ H_+^{N/2})$,
where $H_+= H(r_+)$.  The black hole thermodynamical quantities are
\bea
T &=&\fft{\kappa}{2\pi}\,,\qquad S=\ft14 \Omega_{D-2} H_+^{\fft12 N} r_+^{D-2}\,,\qquad
\Psi=\sqrt{\ft{N q}{q+r_+^{D-3}}}\,,\nn\\
M &=& \ft{\Omega}{16\pi} \Big((D-2) r_+^{D-3} + (D-3) N q\Big)\,,\qquad
Q = \ft{(D-3)\Omega}{16\pi} \sqrt{N q(q + r_+^{D-3})}\,.
\eea
They satisfy the first law $dM=TdS + \Psi dQ$. Here $\Omega$ denotes the volume of the unit $S^{D-2}$.  The solution is extremal when $r_+=0$, corresponding to $M_{\rm ext} = \sqrt{N} Q$. In this limit, $H$ can be any harmonic function of the Euclidean transverse space.

For general non-extremal black holes, we have $M > M_{\rm ext}$. (A particle with $(m,e)$ is a small black hole if $m\ge \sqrt{N} e$.) The entropy depends on the mass and charge, namely
\bea
S &=& \ft14\Omega\, r_+^{D-2} \big(1 - \ft{D-2}{(D-3)N} + \ft{16\pi M}{(D-3)\Omega N r_+
^{D-3}}\big)^{N/2}\,,\nn\\
r_+^{D-3}&=&\ft{8\pi \big[((D-3) N-2 (D-2))M + \sqrt{(D-3)^2 N^2 M^2-4 (D-2) N((D-3) N-D+2)Q^2}\big]}{
(D-2)((D-3)N -D+2)\Omega}\,.
\eea
Although we have not proven analytically the entropy super-additivity for the general case, we have not seen any counter examples in a thorough numerical analysis.  Note that when we split $(M,Q)=(M_1+M_2,Q_1+Q_2)$, the mass-charge bound should be held not only for $(M,Q)$ but also for $(M_1,Q_1)$ and $(M_2,Q_2)$.

We now turn to calculate the gravi-electrostatic force on a charged particle outside the black hole.  Naively, one might simply substitute the metric (\ref{chargedbh}) into the formalism of section 2.  However, there is a subtlety in the EMD theory. The action of the fundamental particle coupled to $A$ should be written in its own particle frame, rather than in the Einstein frame. This is analogous to the string action that should be written in the string frame. (See, e.g.~\cite{Duff:1994an}.) The particle frame can be obtained by the conformal transformation from the Einstein frame $g_{\mu\nu} = e^{a\phi} \hat g_{\mu\nu}$, where we use the hat to denote the quantities in the particle frame. In the particle frame, the Lagrangian becomes
\be
\hat{\cal L}=\sqrt{-\hat g}\, e^{\fft12(D-2)a\phi}\,\Big( \hat R - \ft14 F^2 -\ft12 \big[1- \ft12 (D-1)(D-2) a^2\big]
(\partial\phi)^2\Big)\,,
\ee
and the metric functions of the black hole become
\bea
\hat h = \fft{\tilde f}{H^2}\,,\qquad
\hat f =H^{2-N} \tilde f\,,\qquad
\hat\rho^2 = r^2 H^{N-2}\,.
\eea
The hatted particle action preserves the symmetry $\phi\rightarrow \phi + c$ and $A\rightarrow e^{-\fft12 a c} A$ of the original Lagrangian. Following section 2, we find the equilibrium $r_0$ is determined by
\be
\fft{e}{m} = \ft{2q r_0^{D-3} + (r_0^{D-3}-q) r_+^{D-3}}{2\sqrt{N q r_0^{D-3} (r_0^{D-3}-r_+^{D-3})(r_+^{D-3}+q)}}\,.
\ee
In the extremal $r_+=0$ limit, we have the extremal $e/m=1/\sqrt{N}$, which is independent of $r_0$. This is the manifestation of no force between two extremal black holes.  For general non-extremal black holes with $r_+>0$, we require that $r_0\ge r_+$. The charge/mass ratio is a monotonically decreasing function of $r_0 \in (r_+, \infty)$.  Thus we have
\be
\fft{e}{m}\ge \ft{2q + r_+^{D-3}}{2\sqrt{N q (q+r_+^{D-3})}}\ge \ft{1}{\sqrt{N}}\,.
\ee
In other words, the particle that have a static equilibrium cannot be a black hole.  Two non-extremal black holes carrying the same type of charges always attract, consistent with our earlier observation that they satisfy the entropy super-additivity. The $a=0$ case reproduces the conclusion for the RN black hole \cite{Zhao:2018wkl}. It is worth seeing that equilibria of the particles are all unstable with
\be
\lambda = \big(\ft{r_+}{r_0}\big)^{D-2} \big(\ft{H(r_+)}{H(r_0)}\big)^{N/2}\, \kappa\,,
\ee
satisfying the bound $\lambda <\kappa$, proposed in \cite{Hashimoto:2016dfz}.

\section{EMD theory with two Maxwell fields}

We now generalize section 3 by introducing another Maxwell field.  The Lagrangian in the Einstein frame is
\be
{\cal L}=\sqrt{-g} (R - \ft12 e^{a_1 \phi} F_1^2 - \ft12 e^{a_2\phi} F_2^2)\,,\qquad F_i=dA_i\,,\qquad
a_i^2=\fft{4}{N_i} - \fft{2(D-3)}{D-2}\,.
\ee
For later purpose, we introduce a parameter $\zeta$, defined by
\be
\zeta=-\ft12 a_1 a_2 - \ft{D-3}{D-2}\,.\label{zetadef}
\ee
Singly-charged black holes associated with either $A_1$ or $A_2$ are given in (\ref{chargedbh}) with the fields and constants labeled by subscripts ``1'' or ``2'' appropriately. When $\zeta=0$, exact solutions can be constructed carrying both charges, and the no-force condition exists in the extremal limit, (see, e.g.~\cite{Lu:2013eoa}.) For $\zeta\ne 0$, general analytical solutions are hard to come by.  In $D=4$ and $a_2=-a_1$, approximate solutions were constructed in \cite{Geng:2018jck}. When $a_1 a_2$ is negative, an exact solution always exists for appropriately fixed charges.  To be specific, when
\be
Q_i=\sqrt{-\fft{\epsilon_{ij} a_j}{a_1-a_2}}\, Q\,,
\ee
the solution becomes the RN black hole with charge $Q$. (Here we take $a_1$ and hence $-a_2$ to be positive.)  The two-charge extremal solution is a bound state of two extremal $(Q_1,0)$ and $(0,Q_2)$ black holes, with the binding energy
\be
\Delta M=M_1 + M_2 - M = \sqrt{N_1} Q_1 + \sqrt{N_2} Q_2 - \sqrt{\ft{2(D-2)}{D-3}} Q\,.
\ee
We find that $\Delta M\, (>,=,<)\, 0$ correspond to $\zeta\, (<,=,>)\, 0$ respectively. In string theories, where $\zeta=0$, vanishing $\Delta M$ is a consequence of the no-force condition between the supersymmetric intersecting $p$-branes. The $\Delta M<0$ and $\Delta M>0$ cases were referred to as fission and fission bombs in \cite{lupofission}.  The $\zeta>0$ case with negative binding energy suggests that the extremal black holes are repulsive.  In fact, in $D=4$, and $a_1=-a_2$, for which $\zeta=a_1^2 - 1$, it was demonstrated numerically that entropy super-additivity can be violated for $\zeta>0$ \cite{Geng:2018jck}.  In particular, for $a_1=\sqrt3$, corresponding to the Kaluza-Klein theory, the violation can be established analytically \cite{Cvetic:2018dqf}.

We now calculate the static force between the black holes $(M_1,Q_1)$ and $(M_2,Q_2)$, associated with $A_1$ and $A_2$ respectively, and hence there is no electrostatic force between them. Assuming that $(M_1,Q_1)=(m_1,e_1)$ are sufficiently small to be a point object, we can apply (\ref{noelec}).  In the particle frame associated with $A_1$, the relevant metric functions of the black hole $(M_2,Q_2)$ are
\bea
\hat h = H_2^{\zeta N_2} \tilde f\,,\qquad \hat f = H_2^{-(\zeta+1) N_2} \tilde f\,.
\eea
From the asymptotic behavior of $\hat h$, the antigravity becomes evident for $\zeta>0$.  To be specific, in the $r_+=0$ extremal case, we have no force when $\zeta=0$, for which $h=1$. In general, we find that there is an equilibrium $\hat h'(r_0)=0$, yielding
\be
\big(\fft{r_0}{r_+}\big)^{D-3} =1 + \fft{q_2 + r_+^{D-3}}{\zeta N_2 q_2 -r_+^{D-3}}\,.
\ee
Thus for $\zeta\le 0$, the equilibrium is always inside the horizon, implying that the two objects always attract.  When $\zeta>0$, on the other hand, the equilibrium can be outside the horizon for sufficiently small $r_+$.  In particular, for
$r_+ \in (0, (\zeta N_2 q_2)^{\fft{1}{D-3}})$, we have $r_0 \in (0, \infty)$. In terms of the black hole mass-charge relation, we have
\be
\sqrt{N_2}\le \ft{M_2}{Q_2} \le \ft{\sqrt{N_2}\big( (D-3) + (D-2) \zeta\big)}{(D-3) \sqrt{1+N_2 \zeta}}\,.\label{m2q2cond}
\ee
Note that as the black hole $(M_2,Q_2)$ approaches extremality $r_+\rightarrow 0$, the equilibrium $r_0$ also approaches zero with $r_0/r_+\sim 1 + 1/(\zeta N_2)$. All the equilibria are unstable, with
\bea
\lambda^2 &=& \ft{(D-3)^2\, r_+^{2 (D-3)}\, (\zeta  N_2 q_2 r_+^{3-D}-1)^{\frac{2 (D-2)}{D-3}}}{2 N_2^{N_2+1} q_2^{\frac{2 (D-2)}{D-3}} \zeta ^{N_2+1}\, (\zeta  N_2+1){}^{\frac{D-1}{D-3}-N_2}\, (q_2 r_+^{3-D}+1)^{N_2}}\,.
\eea
(Unlike the previous case, the $\lambda <\kappa$ bound can now be violated.)  Thus for a charged particle $(m_1,e_1)$ outside the black hole $(M_2,Q_2)$, satisfying (\ref{m2q2cond}), there is a potential barrier of the gravi-electrostatic force at certain $r_0>r_+$. As the black hole becomes  extremal, both $r_+$ and $r_0$ approach zero, and barrier becomes infinitesimally thin but infinitely high. The two objects are then repulsive in the whole space. This result is independent of the mass and charge ratio of $(m_1,e_1)$, indicating that the repulsion can exist between two such black holes, mediated by the dilaton.

\section{Conclusions}

We considered a class of EMD theories with two Maxwell fields.  We calculated the static force between one black hole of mass and charge $(M_2,Q_2)$ and the other of $(m_1,e_1)$ that was sufficiently small and could be treated as point-like. It is clear that for large enough mass/charge ratio, black holes will always attract.  However, in the extremal or near extremal cases, the sign of the force depends on the $\zeta$ parameter (\ref{zetadef}) associated with the dilaton coupling constants $(a_1,a_2)$.  For $\zeta<0$, the black holes always attract.  This includes the case when two black holes carrying the same type of charges $(a_1=a_2)$, as demonstrated in section 3. For $\zeta=0$, two black holes in general attract, except when both are extremal, in which case, there is no force between them. This includes black holes that are supersymmetric intersections of $p$-branes. Repulsion occurs when $\zeta>0$ and the black holes are extremal or sufficiently nearly extremal. This reproduces precisely the same result from examining the black hole mass and entropy formulae: the two-charge extremal or near-extremal black holes with $\zeta>0$ are like bound states with negative binding energy and furthermore they can violate the entropy super-additivity.  The two totally different approaches point to the same picture that these black holes will naturally split, like fission bombs. When $a_2=-a_1$, the two electrically-charged black holes can be viewed as dyonic black holes associated with a single field strength.

It should be emphasized that the two black holes of $(M_1,Q_1)$ and $(M_2,Q_2)$ do not have electrostatic force between them, since they are charged under different Maxwell fields. The repulsion is mediated by the dilaton, and our result is indeed sensitive to the parameter $\zeta$. Our conclusion opens up a variety of related questions in black hole physics, such as black hole horizon area and the information paradox when a black hole may break up like a bomb. Since EMD theories are inspired by the strings where the black holes are lower-dimensional artifacts of branes.  The issue of black hole repulsion and related topics should perhaps be addressed by the brane dynamics in strings and/or M-theory.

\section*{Acknolwegement}

We are grateful to Gary Gibbons for discussions. HL and QQZ are supported in part by NSFC grants No.~11875200 and No.~11475024. ZLW is supported in part by NSFC grants No.11305125, No.11447607.

%and the Double First-class University Construction Project of Northwest University.

\end{document}